\documentclass[10pt,letterpaper]{article}
\usepackage{opex3}

\begin{document}

%%%%%%%%%%%%%%%%%% title page information %%%%%%%%%%%%%%%%%%
\title{Spiraling elliptic solitons in nonlocal nonlinear media without anisotropy}

\author{Guo Liang, Qian Shou and Qi Guo$^{*}$}

\address{Laboratory of Photonic Information Technology, South China Normal
University, Guangzhou 510631,China}

\email{guoq@scnu.edu.cn} %% email address is required



\begin{abstract}The optical spatial solitons with ellipse-shaped spots
have generally been considered to be a result of either linear or
nonlinear anisotropy. In this paper, we introduce a class of
spiraling elliptic solitons in the nonlocal nonlinear media without
both linear and nonlinear anisotropy. The spiraling elliptic
solitons carry the orbital angular momentum, which plays a key role
in the formation of such solitons, and are stable for any degree of
nonlocality except the local case when the response function of the
material is Gaussian function. The formation of such solitons can be
attributable to the effective anisotropic  diffraction (linear
anisotropy) resulting from the orbital angular momentum.
Our variational analytical result is confirmed by direct numerical simulation of the nonlocal nonlinear Schr\"{o}dinger
equation.
\end{abstract}

 % REPLACE WITH CORRECT OCIS CODES FOR YOUR ARTICLE

%%%%%%%%%%%%%%%%%%%%%%% References %%%%%%%%%%%%%%%%%%%%%%%%%

%%%%%%%%%%%%%%%%%%%%%%%%%%  body  %%%%%%%%%%%%%%%%%%%%%%%%%%
\section{Introduction}
The nonlinear propagation of optical beams with ellipse-shaped spots
has been discussed during recent years. The self-trapping beams with ellipse-shaped spots can be
obtained by introducing either linear anisotropy or nonlinear
anisotropy. Elliptic incoherent solitons have been reported in
saturable nonlinear media~\cite{Eugenieva-ol-00}, in strongly
nonlocal media~\cite{Shen-oc-07} and in photorefractive screening
nonlinear media~\cite{Katz-ol-04}, where linear anisotropy comes
from the anisotropic coherence function. Furthermore, it was
predicted~\cite{Guo-joa-99} that there exists an elliptical
self-trapping beam for the extraordinary light in uniaxial crystals
due to the anisotropic diffraction (linear
anisotropy)~\cite{Ciattoni-josaa-03,Polyakov-pre-02}. On the other
hand, coherent elliptic strongly nonlocal solitons were observed
experimentally in lead glass~\cite{Rotschild-prl-05} where nonlinear
anisotropy is achieved by rectangular boundaries in the transverse,
and were also simultaneously and independently predicted when the
nonlinear response function of the medium was assumed to be
anisotropic~\cite{Qin-acta-05}. And elliptical discrete solitons can
form in an optically induced two-dimensional photonic lattice where
the nonlinear anisotropy comes of enhanced photorefractive
anisotropy and nonlocality under a nonconventional bias
condition~\cite{Zhang-oe-08}.

Since optical solitons are the result of the exact balance between
linearity and nonlinearity, the elliptic solitons can generally not
exist in the media with both linear and nonlinear isotropy, and the
ellipse-shaped beams always undergo significant oscillations in the
propagation directiom in such
media~\cite{Crosignani-ol-93,Snyder-ol-97,Tichonenko-ol-98}. It was
predicted very recently~\cite{Desyatnikov-prl-10}, however, that the
elliptic solitons with the initial orbital angular momentum (OAM)
can exist in such media, and they will rotate along the propagate
distance.
%They introduced a novel class of spiraling elliptic
%solitons carrying OAM in both local and saturable nonlinear media.
Such solitons are unstable in the (local) cubic nonlinear media, but
they can propagate stably in the saturable nonlinear media because
the saturable nonlinearity can arrest the collapse
instability~\cite{Fibich-jam-99}.

Apart from the saturable nonlinearity, there is another mechanism
that is nonlocal nonlinearity can arrest the catastrophic collapse
of the self-trapping beams~\cite{Bang-pre-02}. So such class of
spiraling elliptic solitons might also exis t in nonlocal nonlinear
media, which will be confirmed theoretically in this paper.
\section{The variational solution of the spiraling elliptic soliton}
The propagation of optical beams in nonlocal cubic nonlinear media
can be modeled by the following nonlocal nonlinear Schr\"{o}dinger
equation(NNLSE)~\cite{Deng-ol-07},
\begin{equation}
2ik\frac{\partial A}{\partial \zeta}+\frac{\partial^2A}{\partial
\xi^2}+\frac{\partial^2A}{\partial
\eta^2}+2k^2\frac{n_2}{n_0}A\int\!\!\!\int
\bar{R}(\xi-\xi',\eta-\eta')|A(\xi',\eta')|^2{\rm d}\xi'{\rm
d}\eta'=0,\label{nnlse in physic system}
\end{equation}
where $A(\xi,\eta,\zeta)$ is a paraxial beam, $\bar{R}$ is the
response function of the medium, $\zeta$ is the longitudinal
coordinate, $\xi$ and $\eta$ are the transverse coordinates,
$k=\omega n_0/c$ is the wavenumber in the media without
nonlinearity, $n_0$ is the linear refractive index of the media,
$n_2$ is the nonlinear index coefficient. Through the dimensionless
transformation $
x=\xi/w_0,y=\eta/w_0,z=\zeta/(kw_0^2),\psi=Akw_0(n_2/n_0)^{1/2},R=w_0^2\bar{R},$
where $w_0$ is the initial width of the optical beam, Eq.(\ref{nnlse
in physic system}) is expressed as in the dimensionless form
\begin{equation}
i\frac{\partial \psi}{\partial
z}+\frac{1}{2}\left(\frac{\partial^2\psi}{\partial
x^2}+\frac{\partial^2\psi}{\partial y^2}\right)+\Delta
n\psi=0,\label{nnlse}
\end{equation}
where $\Delta n=\int\!\!\!\int
R(\textbf{r}-\textbf{r}')|\psi(\textbf{r}')|^2{\rm d}^2\textbf{r}'$
is the nonlinear perturbation of refraction index with
$\textbf{r}=x\textbf{e}_x+y\textbf{e}_y$. $R(r)=1/(2\pi
w_m^2)\exp[-r^2/2w_m^2]$ is assumed in this paper, where $w_m$ is the characteristic length of
the response function in the dimensionless system.

The Lagrangian of Eq.(\ref{nnlse}) can be expressed
as~\cite{Guo-oc-06} $L=1/2\int\!\!\!\int(\psi^*\partial
\psi/\partial z-\psi\partial \psi^*/\partial z){\rm d}x{\rm d}y-H, $
where $H$ is the Hamiltonian of this system,
$H=1/2\int\!\!\!\int(\left|\partial\psi/\partial
x\right|^2+\left|\partial\psi/\partial y\right|^2-\Delta
n|\psi|^2){\rm d}x{\rm d}y. $ We introduce a trial
function~\cite{Desyatnikov-prl-10},
\begin{equation}
\psi(x,y,z)=\sqrt\frac{P}{\pi
b(z)c(z)}G\left[\frac{X}{b(z)}\right]G\left[\frac{Y}{c(z)}\right]\exp(i\phi),\label{trial
function}
\end{equation}
where the Gaussian envelope is $G(t)=\exp(-t^2/2)$, the phase is
$\phi=B(z)X^2+\Theta(z)XY+Q(z)Y^2+\varphi(z)$, $X=x\cos
\beta(z)+y\sin\beta(z),Y=-x\sin\beta(z)+y\cos\beta(z)$ and $P$ is
the power, $P=\int\!\!\!\int|\psi|^2{\rm d}x{\rm d}y$. We can obtain
the orbital angular momentum(OAM),
$M=\mbox{Im}\int\!\!\!\int\psi^*(\textbf{r}\times\nabla\psi){\rm
d}^2\textbf{r}=1/2P(b^2-c^2)\Theta$. Inserting the Gaussian ansatz
(\ref{trial function}) into the Lagrangian, $L$ can be analytically
determined. Then using the variational approach, we can obtain that
$P'=0, H'=0, M'=0, b'=2bB, c'=2cQ, \beta
'=(b^2+c^2)\Theta/(b^2-c^2)$ and
\begin{equation}
\varphi'=-\frac{b^2+c^2}{2b^2c^2}+\frac{P[6b^2c^2+5w_m^2(b^2+c^2)+4w_m^4]}{8\pi[(b^2+w_m^2)(c^2+w_m^2)]^{3/2}}\label{phi
prime},
\end{equation}
where the primes indicate derivatives with respect to the variable
$z$. So it can be found that the power, the Hamiltonian and the OAM
of the system are conservative. We can determine the Hamiltonian of
the system, $H=P/4(b'^2+c'^2+\Pi),$ where
{\setlength\arraycolsep{2pt}
\begin{equation}
\Pi=\frac{1}{b^2}+\frac{1}{c^2}+\frac{4 b^2 \sigma
^2}{\left(b^2-c^2\right)^2}+\frac{4 c^2 \sigma
^2}{\left(b^2-c^2\right)^2}-\frac{P}{\pi
\sqrt{\left(b^2+w_m^2\right)
\left(c^2+w_m^2\right)}}\label{potential function},
\end{equation}}
with $\sigma\equiv M/P=1/2(b^2-c^2)\Theta$.

\begin{figure}[htb]
\centerline{\includegraphics[width=9cm]{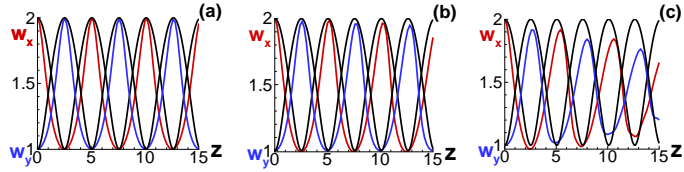}}
\caption{(color online) Comparison of the beam width of the
analytical solution (black curves) with that of the numerical simulation (red
curves for $w_x$ and blue curves for $w_y$) for $\delta=7.5$ (a),
$\delta=4$ (b), and $\delta=2$ (c).}\label{beamwidth}
\end{figure}

\begin{figure}[htb]
\centerline{\includegraphics[width=7cm]{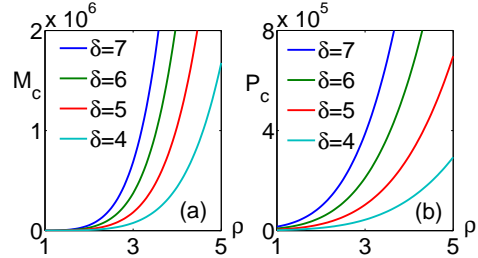}}
\caption{(color online) Critical OAM (a) and critical power (b) as
functions of the ellipticity $\rho$ for different degree of
nonlocality.}\label{relations}
\end{figure}

Solitons can be found as the extrema of the potential $\Pi(b,c)$. Assuming $b>c$ without loss
of generality and letting $\partial\Pi/\partial b=0$ and $\partial\Pi/\partial c=0$,
we can obtain the critical power and the critical OAM
%{\setlength\arraycolsep{2pt}
\begin{eqnarray}
P_c=\frac{2\pi\left(1+\rho^2\right)^3\left[\left(1+\delta^2\right)\left(1+\delta^2\rho^2\right)\right]^{3/2}}{\rho\left[1+\left(6+4\delta^2\right)\rho^2+\left(1+4\delta^2\right)\rho^4\right]}\label{critical
power},\sigma_c=\frac{\left(\rho^2-1\right)^2\left[1+\delta^2\left(1+\rho^2\right)\right]^{1/2}}{2\rho[1+\left(6+4\delta^2\right)\rho^2+\left(1+4\delta
^2\right)\rho^4]^{1/2}},\end{eqnarray}and $M_c=P_c\sigma_c$, where $\rho=b/c$ and $\delta=w_m/b$
represent the ellipticity of the elliptic beam and the degree of
nonlocality, respectively. The larger is $\delta$, the stronger is
the degree of nonlocality. When $P=P_c,\sigma=\sigma_c$, the optical beam can propagate keeping its elliptic profile changeless and
rotating stably. We can also obtain the rotation velocity
$\Omega_c\equiv\beta'=2 (b^2+c^2)\sigma_c/(b^2-c^2)^2$. When the
semi-axes $b$ and $c$ are given, the critical power and the critical
OAM of the spiraling elliptic solitons can be determined by
Eq.({\ref{critical power}}). One example is shown in
Fig.\ref{beamwidth}(a) with $P_c=1.27\times10^5 , \sigma_c=0.561,
\Theta_c=2\sigma_c/(b^2-c^2)=0.374$ and $\Omega_c=0.623$ when
$b=2.0,c=1.0, w_m=15.0$. Comparing two half widths obtained from
variational solution, $w_x=(b^2 \cos^2\Omega_c z+c^2 \sin^2\Omega_c
z)^{1/2}$ and $w_y=(c^2 \cos^2\Omega_c z+b^2 \sin^2\Omega_c
z)^{1/2}$, with those from the numerical simulation of Eq.(\ref{nnlse}) by
using $\psi(x,y,0)=[P_c/(\pi
bc)]^{1/2}\exp[-x^2/(2b^2)-y^2/(2c^2)]\exp(i\Theta_c xy)$ as the
input beam at $z=0$, we find an excellent agreement as shown in
Fig.\ref{beamwidth}(a). The formation of the spiraling elliptic
soliton is due to the effective anisotropic diffraction resulting
from the OAM, which will be illustrated in the fourth part of the
paper, then the ellipticity $\rho$ of the elliptic beam should
increase when the critical OAM $M_c$ increases, as shown in
Fig.\ref{relations}(a). In addition, the OAM can strengthen
effectively diffraction against self-focusing
~\cite{Desyatnikov-prl-10}, so the critical power $P_c$ should
increase together with $M_c$ when $\rho$ increases, as shown in
Fig.\ref{relations}(b). Besides, $P_c$ and $M_c$ increase when the
degree of nonlocality increases, which can also be observed in Fig.~\ref{relations}.

It can be shown that the solution Eq.(\ref{trial function}) in the
strong nonlocality is equivalent to the Gaussian
complex-variable-function(CVF)-Gaussian soliton, a special case of
the CVF-Gaussian solitons suggested recently~\cite{Deng-ol-09} when
an arbitrary analytical function $f$ takes the Gaussian function.
For the limit of the strongly nonlocal nonlinearity, Eq.({\ref{phi
prime}}) and Eq.({\ref{critical power}}) can be reduced as $\varphi
'|_{w_m\rightarrow \infty}=-(b^2+c^2)/(2b^2c^2),P_c|_{w_m\rightarrow
\infty}=\pi w_m^4(b^2+c^2)^2/(2b^4c^4),\sigma_c|_{w_m\rightarrow
\infty}=(b^2-c^2)^2/(4b^2c^2)$. If we use the variable substitutions
$b^2=\kappa^2w^2/(\kappa^2+1), c^2=\kappa^2w^2/(\kappa^2-1)$, the
spiraling elliptic soliton~(\ref{trial function}) can be deduced as
%\begin{equation}\label{G-CVF-soliton}
$\psi(x,y,z)=\sqrt{{P}/({\pi
bc})}f(\upsilon)\exp\left[-{r^2}/({2w^2})-i\beta\right]$,
%\end{equation}
where$f(\upsilon)=\exp(-\upsilon^2/2)$ and $\upsilon=(x+iy)/(\kappa
w)\exp(-i\beta)$. The expression does be the Gaussian CVF-Gaussian soliton~\cite{Deng-ol-09}, and the
parameter $\kappa$ here is the distribution factor $b$ in
Ref.~\cite{Deng-ol-09}.
\section{Analytic stability analysis of the solution }
From Eq.(\ref{critical power}), we know that $P_c$ and $\sigma_c$
can be determined when b and c are given. It is also true in
reverse. When $P_c$ and $\sigma_c$ are given first, then $b$ and $c$
can be obtained, which are corresponding to the stationary point of
the potential function $\Pi(b,c)$. We use $(b_s,c_s)$ to represent
this stationary point here. Hence we can study the stability of our
analytical soliton solution by determining whether $(b_s,c_s)$ is
the minimum point of $\Pi(b,c)$ or not. For this purpose, we expand
$\Pi(b,c)$ in Taylor's series about the
stationary point $(b_s,c_s)$ to the second order~\cite{Riley-book-2002}
\begin{equation}
\Pi(b,c)\approx\Pi(b_s,c_s)+\frac{1}{2}\left[\mu_1(\Delta
b+\frac{\mu_2}{\mu_1}\Delta
c)^2+\frac{\mu_1\mu_3-\mu_2^2}{\mu_1}(\Delta
c)^2\right],\label{potential expand}
\end{equation}
where $\mu_1=\partial^2\Pi/\partial
b^2\left|_{P=P_c,\sigma=\sigma_c}\right.$, $
\mu_2=\partial^2\Pi/(\partial b\partial
c)\left|_{P=P_c,\sigma=\sigma_c}\right.$, $
\mu_3=\partial^2\Pi/\partial
c^2\left|_{P=P_c,\sigma=\sigma_c}\right.$, $\Delta b=b-b_s$ and $\Delta
c=c-c_s$. We can derive that $\mu_1, \mu_3>0$ and $\mu_2^2<\mu_1\mu_3$
when $w_m\neq0$. So the stationary point $(b_s,c_s)$ is really the
minimum point of the potential function because
$\Pi(b,c)-\Pi(b_s,c_s)>0$ in this case. However, we obtain that
$\mu_2^2=\mu_1\mu_3$ when $w_m=0$, and Eq.(\ref{potential expand}) is deduced as $\Pi(b,c)-\Pi(b_s,c_s)\approx1/2[\mu_1(\Delta
b+\mu_2\Delta c/\mu_1)^2]$. Thus $\Pi(b,c)-\Pi(b_s,c_s)\approx0$ along the particular direction
that $\Delta b=-\mu_2\Delta c/\mu_1$, and next higher order term need to be considered in order to judge whether $(b_s,c_s)$ is minimum point.
% Then the Taylor expansion has to be taken to a higher order.
But we can deal with the problem in a simpler way. We directly compare the value of the potential
function at the stationary point $(b_s,c_s)$ and that at the
point with a displacement $(\Delta b,\Delta c)$ from the stationary
point along the particular direction that $\Delta b=-\mu_2\Delta
c/\mu_1$, and find that $\Pi(b_s+\Delta b,c_s+\Delta
c)-\Pi(b_s,c_s)=0$. Therefore, for the case of $w_m=0$ the
stationary point $(b_s,c_s)$ is not the minimum point of the
potential function yet. As a result, we can draw the conclusion that
the soliton solutions are stable for any degree of nonlocality
except for the local case.

It is known that the spatial profile
of the optical beams in the local cubic nonlinear media will evolve to a specific circularly symmetric shape,
known as the Townes profile~\cite{Moll-prl-03} that is very
different from Gaussian profile. So when the degree of nonlocality
is weak enough ($\delta$ is small enough), the Gaussian trial function is not suitable any longer. The deviation between the exact solution
of Eq.(\ref{nnlse}) and the trial solution Eq.(\ref{trial function})
can bring about the disagreement of the beam width obtained from
the variational approach with that from the numerical simulation, as shown in
Fig.\ref{beamwidth}(c) when $\delta=2$.
In fact, the deviation between the
variational solution and the numerical simulation  was
discussed in Ref.~\cite{Desyatnikov-prl-10} for the local case.
%, as shown in Fig.3(c) and Fig.3(f) ofRef.~\cite{Desyatnikov-prl-10}.
% When the degree of nonlocality is weak
%enough, the variational approach with the Gaussian trial function does not work any longer.
%, because
%we can not find a suitable analytical trial function expressed in
%terms of elementary functions.
But, even so, we can find that when $\delta=4$, which is out of the region
of strong nonlocality, our variational results still have a
good agreement with the numerical simulations, as shown in
Fig.\ref{beamwidth}(b).
\section{Physical explanation of the formation of spiraling elliptic solitons}
To better understand the formation of such spiraling elliptic
solitons, we turn to the analysis of the wave vector of the electric
field that can be obtained by $\textbf{k}=\nabla \bar{\phi}_{tol}$,
where $\bar{\phi}_{tol}$ is the total phase of the electric field
 expressed as
$\bar{\phi}_{tol}(\xi,\eta,\zeta)=\bar{\phi}(\xi,\eta,\zeta)+k\zeta$
in the physical coordinate system. Taking the paraxial beam into
consideration, we need only take care of the wave vector \textbf{k}
around some point on the propagation axis ($0,0,\zeta_0$). We
therefore can expand $\bar{\phi}(\xi,\eta,\zeta)$ with respect to
$(\xi,\eta,\zeta)$ in Taylor's series about ($0,0,\zeta_0$) to the
second order, and obtain
%\begin{eqnarray}
%\textbf{k}&=&[\gamma_\xi+\gamma_{\xi\xi}\xi+\gamma_{\xi \eta}\eta+\gamma_{\xi \zeta}(\zeta-\zeta_0)]\textbf{e}_\xi+[\gamma_\eta+\gamma_{\eta\eta}\eta+\gamma_{\xi \eta}\xi+\gamma_{\eta\zeta}(\zeta-\zeta_0)]\textbf{e}_\eta\nonumber\\
%&+&[\gamma_\zeta+\gamma_{\zeta\zeta}(\zeta-\zeta_0)+\gamma_{\eta\zeta}\eta+\gamma_{\xi
%\zeta}\xi+k]\textbf{e}_\zeta,
%\end{eqnarray}}
\begin{equation}
\textbf{k}=(\gamma_{\xi\xi}\xi+\gamma_{\xi\eta}\eta)\textbf{e}_\xi+(\gamma_{\eta\eta}\eta+\gamma_{\xi\eta}\xi)\textbf{e}_\eta+
%\gamma_\xi\textbf{e}_\xi+\gamma_\eta\textbf{e}_\eta+
k\textbf{e}_\zeta\label{k expand},
\end{equation}where
$\gamma_j=\partial_j\bar{\phi}|_{\xi=0,\eta=0,\zeta=\zeta_0},\gamma_{jl}=\partial^2_{jl}\bar{\phi}|_{\xi=0,\eta=0,\zeta=\zeta_0}$
($j,l=\xi,\eta,\zeta$). In the equation above, we neglect the terms
$\gamma_{\xi \zeta}$, $\gamma_{\eta \zeta}$, $\gamma_{\zeta \zeta}$
and $\gamma_{\zeta}$ because of the fact~\cite{Chi-ol-95} that
$\partial \bar{\phi}/\partial \zeta\ll\partial \bar{\phi}/\partial
\xi (\mathrm{or}~\partial \bar{\phi}/\partial \eta)$ for paraxial
beams, and take $\gamma_\xi, \gamma_\eta=0$ (if $\gamma_\xi,
\gamma_\eta\neq0$, the wave vector would have an inclination angle
with respect to the $\zeta$-axis).

\begin{figure}[htb]
\centerline{\includegraphics[width=6cm]{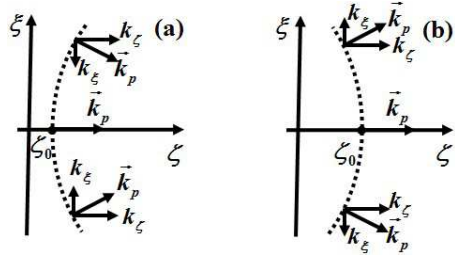}}
\caption{Schematic of the vector $\mathbf k_p$ %about the on-axis point $(0,0,\zeta_0)$
at the position ($\xi,0,\zeta$) for $\gamma_{\xi\xi}<0$ (a) and
$\gamma_{\xi\xi}>0$ (b). The points we consider are in fact very close
to the point ($0,0,\zeta_0$), but we have magnify the extent just
for the sake of greater clarity. }\label{wavevector}
\end{figure}

Equation~(\ref{k expand}) tells the fact that the pointing of the vector \textbf k at the position $(\xi,\eta,\zeta)$ depends upon the sign of $\gamma_{\xi\xi}$, $\gamma_{\eta\eta}$, and $\gamma_{\xi\eta}$.
For simpleness, we take the projection of \textbf k on the ($\xi,0,\zeta$)-plane, representing by $\mathbf k_{p}$,
%$\xi$-component and $\zeta$-component of \textbf k
into
consideration, and the situation for the projection of \textbf k on the ($0, \eta,\zeta$)-plane can be
dealt with in the same way. First at the position $(\xi,0,\zeta)$,
$k_\xi=\gamma_{\xi\xi}\xi$. If $\gamma_{\xi\xi}<0$, we can reach a
conclusion that $k_\xi<0$ in the upper half plane $(\xi>0)$ and
$k_\xi>0$ in the lower half plane $(\xi<0)$. Then $\textbf{k}_p$
points downward in the upper half plane and upward in the lower half
plane, as shown in Fig.\ref{wavevector}(a). As a result, the optical beam will be contracted along the $\xi$ direction when
$\gamma_{\xi\xi}<0$. When $\gamma_{\xi\xi}>0$, on the
contrary, $\textbf{k}_p$
points upward in the upper half plane and downward in the lower half
plane,
as shown in Fig.\ref{wavevector}(b), and the optical beam will be expanded along the $\xi$ direction. For the situation of the position ($\xi,\eta(\neq0),\zeta)$, although the presence of the cross-talking term $\gamma_{\xi\eta}$ makes it somewhat complicated, the pointing of $\textbf k_p$ can be determined in the similar way.

The discussion above about the physical mechanism is based on the analogy of the optical beam %in the space domain
and the
optical pulse. % in the time domain. %It is well-known
%that the group-velocity dispersion-induced temporal
%effects[14] for the pulse and the diffraction-induced
%spatial effects[13] for the beam are a closely analogous
%pair. Moreover,
Both of them can be dealt with in
Fourier analysis---the temporal frequency of the pulse
is the analogue of the spatial spectrum of the beam. Therefore, by analogy with the phenomenon of the chirp, the time dependence of the phase for the optical pulse~\cite{Agrawal-book-2001}, the transverse-space dependence of the phase for the optical beam can be referred to as ``spatial chirp'', and the first four terms in Eq.~(\ref{k expand}) are linear spatial chirp terms (the similar concept has been introduced in Ref.~\cite{Faccio-JOSAB-2005}). As a result, the physical mechanism for the broadening (shortening) of the optical pulse~\cite{Agrawal-book-2001} and the expanding (contracting) of the optical beam can be understood in this uniform sense.

 On that basis, we discuss the part of the phase due
to the OAM expressed as
$\bar{\phi}_{OAM}=M\sin2k\beta(-\xi^2+\eta^2)/[P(b^2-c^2)]+2\xi\eta
M\cos2k\beta/[P(b^2-c^2)] $ in the physical coordinate system, which
is corresponding to $\phi_{OAM}=\Theta XY$ in the dimensionless
coordinate system. Then we can obtain the part of the wave vector
%in $\xi$-direction and $\eta$-direction respectively
caused by the OAM
\begin{equation}
k_\xi^{(OAM)}=-\frac{2M\sin2k\beta}{P(b^2-c^2)}\xi+\frac{2M\cos2k\beta}{P(b^2-c^2)}\eta,
k_\eta^{(OAM)}=\frac{2M\sin2k\beta}{P(b^2-c^2)}\eta+\frac{2M\cos2k\beta}{P(b^2-c^2)}\xi\label{wave
vector},
\end{equation}
where
$k_\xi^{(OAM)}=\partial_\xi\bar{\phi}_{OAM},k_\eta^{(OAM)}=\partial_\eta\bar{\phi}_{OAM}$.
%The first terms in Eq.(\ref{wave vector}) can lead to the broadening
%or compression of the beam-spot depending on the coefficients before
%$\xi$ and $\eta$. The positive coefficients make a beam broaden and
%the negative coefficients make it compress.
From Eq.(\ref{wave vector}) we can find that the contributions of
OAM to the wavevector are different (asymmetric) in $\xi$-direction and
$\eta$-direction, because the signs of the two first
terms in Eq.~(\ref{wave vector}) are
opposite. In other words, OAM can result in an effective anisotropic
diffraction. It is the effective anisotropic diffraction that leads
to the formation of elliptic solitons.
\section{Conclusion}
We have obtained spiraling elliptic solitons in nonlocal nonlinear
media without anisotropy by use of the variational approach. The
formation of such solitons is due to an effective anisotropic
diffraction resulting from the orbital angular momentum. We show
that this class of solitons are stable for any degree of nonlocality
except the local case. Our approximate analytical results have been confirmed
by direct numerical simulations of the NNLSE.
\section*{Acknowledgments}
This research was supported by the National Natural Science
 Foundation of China (Grant Nos. 11074080 and 10904041), the Specialized Research Fund for the Doctoral Program
 of Higher Education (Grant No. 20094407110008), and the Natural Science Foundation
of Guangdong Province of China (Grant No. 10151063101000017).

\begin{thebibliography}{99}

\bibitem{Eugenieva-ol-00} E. D. Eugenieva and D. N. Christodoulides, ``Elliptic incoherent solitons in saturable nonlinear media,'' Opt. Lett. \textbf{25,}
972--974 (2000).
\bibitem{Shen-oc-07} M. Shen, Q. Wang and J. L. Shi, ``Elliptic incoherent accessible solitons in strongly nonlocal media,'' Opt. Commun. \textbf{270,}
384--390 (2007).
\bibitem{Katz-ol-04} O. Katz, T. Carmon, T. Schwartz, M. Segev and D. N. Christodoulides, ``Observation of elliptic incoherent spatial solitons,'' Opt. Lett. \textbf{29,}
1248--1250 (2004).
\bibitem{Guo-joa-99} Q. Guo and S. Chi, ``Nonlinear light beam propagation in
uniaxial crystals: nonlinear refractive index, self-trapping and
self-focusing,'' J. Opt. A: Pure Appl. Opt.\textbf{2} 5--15 (2000).
\bibitem{Ciattoni-josaa-03} A. Ciattoni and C. Palma, ``
Optical propagation in uniaxial crystals orthogonal to the optical
axis: paraxial theory and beyond,'' J. Opt. Soc. Am. A, \textbf{20,}
2163--2171 (2003).
\bibitem{Polyakov-pre-02} S. V. Polyakov and G. I. Stegeman, ``Existence and properties of quadratic solitons in anisotropic media:
Variational approach,'' Phys. Rev. E, \textbf{66,} 046622 (2002).
\bibitem{Rotschild-prl-05} C. Rotschild, O. Cohen, O. Manela and M. Segev, ``Solitons in Nonlinear Media with an Infinite Range of Nonlocality: First Observation
of Coherent Elliptic Solitons and of Vortex-Ring Solitons,'' Phys.
Rev. Lett \textbf{95} 213904 (2005).
\bibitem{Qin-acta-05} X. J. Qin, Q. Guo, W. Hu and S. Lan, ``Strongly nonlocal elliptical spatial optical soliton,''  Acta Phys. Sin. \textbf{55,}
1237--1243 (2006)(in Chinese).
\bibitem{Zhang-oe-08} P. Zhang et al., ``Elliptical discrete solitons supported by
enhanced photorefractive anisotropy,'' Opt. Express. \textbf{16,}
3865--3870 (2008).
\bibitem{Crosignani-ol-93} B. Crosignani and P. Di Porto, ``Nonlinear propagation in Kerr media of beams with unequal transverse widths,'' Opt. Lett. \textbf{18,} 1394--1396
(1993).
\bibitem{Snyder-ol-97} A. W. Snyder and D. J.
Mitchell, ``Mighty morphing spatial solitons and bullets,'' Opt.
Lett. \textbf{22,} 16--18 (1997).
\bibitem{Tichonenko-ol-98} V. Tichonenko, ``Observation of
mighty morphing spatial solitons,'' Opt. Lett. \textbf{23,} 594--596
(1998).
\bibitem{Desyatnikov-prl-10} A.S.Desyatnikov, D.Buccoliero, M.R.Dennis and Y.S.Kivshar, ``Suppression of Collapse for Spiraling Elliptic Solitons,'' Phys. Rev. Lett \textbf{104}
053902 (2010).
\bibitem{Fibich-jam-99} G. Fibich and G. Papanicolaou, ``Self-focusing in the perturbed and unperturbed nonlinear Schr$\ddot{o}$dinger equation in critical
dimension, '' SIAM J. Appl. Math. \textbf{60,} 183 (1999).
\bibitem{Bang-pre-02} O. Bang et al., ``Collapse arrest and soliton stabilization in nonlocal nonlinear media,'' Phys. Rev. E \textbf{66,}
046619 (2002).
\bibitem{Deng-ol-07} D. M. Deng and Q. Guo, ``Ince-Gaussian solitons in strongly nonlocal
nonlinear media,'' Opt. Lett. \textbf{32,} 3206--3208 (2007).
\bibitem{Guo-oc-06} Q. Guo, B. Luo, S. Chi, ``Optical beams in sub-strongly non-local nonlinear media:
A variational solution,'' Opt. Commun. \textbf{259,} 336--341
(2006).
\bibitem{Deng-ol-09} D. M. Deng, Q. Guo and W. Hu, ``Complex-variable-function¨CGaussian solitons,'' Opt. Lett. \textbf{34,}
43--45 (2009).
\bibitem{Riley-book-2002} K. F. Riley, M. P. Hobson and S. J. Bence, \emph{Mathematical Methods for Physics and Engineering},
(Cambridge, 2nd ed, 2002), pp. 165-170.
\bibitem{Moll-prl-03} K. D. Moll, A. L. Gaeta and G. Fibich, ``Self-Similar Optical Wave Collapse: Observation of the Townes Profile,'' Phys. Rev. Lett \textbf{90}
203902 (2003).
\bibitem{Chi-ol-95} S. Chi and Q. Guo, ``Vector theory of self-focusing of an optical beam in Kerr media,'' Opt. Lett. \textbf{20,}
1598--1600 (1995).
\bibitem{Agrawal-book-2001}G. P. Agrawal, {\em Nonlinear Fiber Optics} (Academic, 3rd ed, San Diego, 2001), pp. 63-134.
\bibitem{Faccio-JOSAB-2005} D. Faccio, P. D. Trapani, S. Minardi and A. Bramati, ``Far-field spectral characterization of conical
emission and filamentation in Kerr media,'' J. Opt. Soc. Am. B,
\textbf{22,} 862--869 (2005).
\end{thebibliography}
\end{document}